\documentclass[10pt,letterpaper]{article}
\usepackage{amsfonts, latexsym, amsmath}
\usepackage{bbm}
\usepackage[latin1]{inputenc}
\usepackage[english]{babel}

\usepackage{graphicx}
\usepackage{url}
\usepackage{subfigure}
\usepackage{psfrag}
\usepackage{multirow}

\newtheorem{defin}{Definition}
\newtheorem{prop}{Proposition}
\newtheorem{lemma}{Lemma}

\newtheorem{remark}{Remark}
\newtheorem{ex}{Example}
\newtheorem{prob}{Problem}
\newtheorem{cond}{Condition}
\newcommand{\Enc}{\mathsf{Enc}}
\newcommand{\Dec}{\mathsf{Dec}}
\newcommand{\R}{\mathbb{R}}

\newcommand{\demo}{{\em Proof.}\,}
\newcommand{\findemo}{\hfill$\Box$}

\newcommand{\highlight}[1]{\textbf{#1}}
\newcommand{\discretAB}[2]{\{ #1, \ldots, #2 \}}

\newcommand{\lar}{\leftarrow}

\newcommand{\Pp}[1]{\Pr \left [ #1 \right ]}
\newcommand{\Ppp}[2]{\Pr_{#1}\left [ #2 \right ]}

\newcommand{\X}{\mathcal{X}}
\newcommand{\Y}{\mathcal{Y}}
\newcommand{\Se}{\mathcal{S}}
\newcommand{\DB}{\mathcal{DB}}

\title{Identification with Encrypted Biometric Data\thanks{An extended abstract -- entitled ``Error-Tolerant Searchable Encryption'' -- of this work has been accepted to and will be presented at the Communication and Information Systems Security Symposium, International Conference on Communications (ICC) 2009, June 14-18, Dresden, Germany. This paper ``Identification with Encrypted Biometric Data'' is the full version of our work.}}

\author{Julien Bringer$^1$, Hervé Chabanne$^{1,2}$ and Bruno Kindarji$^{1,2}$
\thanks{This work was partially supported by the french ANR RNRT project BACH.} \\
$^1$ Sagem S\'ecurit\'e, Osny, France.\\
$^2$ Institut TELECOM, Télécom ParisTech, Paris, France.}

\begin{document}

\date{}
\maketitle
\pagestyle{empty}

\begin{abstract}
	Biometrics make human identification possible with a sample of a biometric trait and an associated database. Classical identification techniques lead to privacy concerns. This paper introduces a new method to identify someone using his biometrics in an encrypted way.

Our construction combines Bloom Filters with Storage and Locality-Sensitive Hashing. We apply this error-tolerant scheme, in a Hamming space, to achieve biometric identification in an efficient way. This is the first non-trivial identification scheme dealing with fuzziness and encrypted data. 

\end{abstract}

\section{Introduction}

The arising of biometric recognition systems is based on the uniqueness of some natural information every human being carries along. For instance, it is possible to verify that a given individual is the one he claims to be (\emph{Verification}). It is also possible to find someone's identity among a collection thanks to his biometrics (\emph{Identification}).

In this paper, we design a biometric identification system that is based on encrypted data, so that privacy is guaranteed, and in a way that does not take too much time and memory to process. For that purpose, we need to find a way to: 
\begin{itemize}
\item mitigate the effects of biometrics fuzziness, 
\item and efficiently identify someone over an encrypted database. 
\end{itemize}
It follows the idea of searchable encryption and we here explain how to make efficient queries to the database, that look for a pattern close to a given one in encrypted data, \textit{i.e.} a search with error-tolerance.

\subsection{Related Works and Motivation}
Security of biometric systems is widely studied -- cf. \cite{JaNaNa08,JaRoUl05,BoCoRa02}
 -- and although a lot of vulnerabilities are now well understood and controlled, it is still difficult to achieve an end-to-end system which satisfies all constraints. In particular, biometric template privacy is an important issue due to the non-revocability and non-renewability of biometric features.

\subsubsection{\noindent {\bf Biometrics and Cryptography} }

	A specific difficulty concerning biometrics is their fuzziness. It is nearly impossible for a sensor to obtain the same image from a biometric data twice: there will always be significant differences. The classical way to supersede variations between different captures is to use a matching function, which basically tells if two measures represent the same biometric data or not.
	
The integration of biometrics into cryptographic protocols is thus difficult as state-of-the-art protocols are not designed for error-tolerance and fuzziness in their inputs. The two main leads for that are achieving a good stable coding of the data or  making the matching algorithm part of the protocol.
	
	Both sides of the problem are quite hard. The extraction of a constant-length vector has been studied for the iris \cite{Daugman93} and the fingerprint \cite{JainPHP99,TuylsAKSBV05}; the result is a fixed-length bit string on which the matching is realized with the Hamming distance. Following this, we solely focus in this paper on binary biometric data compared with Hamming distance. 
	
	Most of protocols involving biometric data and cryptography use Secure Sketches or Fuzzy Extractors \cite{DodisRS04,JuelsW99}. It uses error correction to reduce variations between the different measures, and to somehow hide the biometric data behind a random codeword -- e.g. \cite{SuLiMe07,NaNaJa07,HaAnDa06,BrChDo06,BringCCKZ07,BoBW07}. 
		
	On the other hand, several biometrics \emph{verification} protocols, \textit{e.g.} \cite{BrCT07,DBLP:conf/acisp/BringerCIPTZ07,DBLP:conf/cans/BringerCPT07,SchoenmakersT06,DBLP:conf/ispec/TangBCP08}, have proposed to embed the matching directly. They use the property of homomorphic encryption schemes to compute the Hamming distance between two encrypted templates. 	Some other interesting solutions based on adaptation of known cryptographic protocols are also investigated in \cite{DBLP:conf/africacrypt/BringerC08,DBLP:conf/iwsec/BringerCPZ08}.
	
	The drawback with all these techniques is that they do not fit well with \emph{identification} in large databases as the way to run an identification among $N$ data would be to run almost as many authentication algorithms. As far as we know, no non-trivial protocol for biometric \emph{identification} involving privacy and confidentiality features has been proposed yet.
	
\subsubsection{\noindent {\bf Identification}} 
	Several algorithms have been proposed for the so-called \emph{Nearest Neighbour} and \emph{Approximate Nearest Neighbour} (\highlight{ANN}) problems. Indyk wrote a review on these topics in \cite{indyk04nearest}. 
	 Recently, Hao~\emph{et al.} \cite{Hao07} demonstrated the efficiency of the \highlight{ANN} approach for iris biometrics where projected values of iris templates are used to speed up identification requests into a large database;  indeed \cite{Hao07} derived a specific \highlight{ANN} algorithm from the iris structure and statistical properties. However, in their construction the iris biometric data are never encrypted, and the way they boost the search for the nearest match reveals a large amount of information about sensitive data.
		
	Our works are also influenced by the problem of finding a match on encrypted data. 
	 Boneh~\emph{et al.} defined the notion of \emph{Public-key encryption with Keyword Search} (\highlight {PEKS}) \cite{BonehCOP04}, in which specific trapdoors are created for the lookup of keywords over public-key encrypted messages. Several other papers, e.g.  \cite{goh03,BethencourtSW06,ByunLL06,Khader06,RyuT07}, have also elaborated solutions in this field. 
	 	 However the main difference between the search for a keyword as understood by Boneh~\emph{et al.}~\cite{BonehCOP04,BonehKOS07} and biometric matching is that an exact match for a given bit string in the plaintext suffices for the former, but not for our motivation. For this purpose, we introduce  a new model for error-tolerant search in Sec. \ref{sec:model} and specific functions to take into account fuzziness in Sec. \ref{sec:def:lsh}. 

The most significant difference here from the primitives introduced previously in \cite{BonehCOP04} is that messages are no longer associated to keywords. Moreover, our primitives enable some imprecision on the message that is looked up. For example, one can imagine a mailing application, where all the mails are encrypted, and where it is possible to make queries on the mail subject. If there is a typo in the query, then looking for the correct word should also give the mail among the results -- at least, we would like that to happen. Note that wildcards are not well-adapted to this kind of application, as a wildcard permits to catch errors providing that we know where it is located, whereas error-tolerance does not have this constraint.

\subsection{Construction Outline}

We propose to use recent advances done in the fields of similarity searching and public-key cryptography. Our technique narrows our identification to a few candidates. In a further step, we must complete it by fine-tuning the results in checking the remaining identities so that the identification request gets a definite answer.

The first step is accomplished by combining Bloom filters with locality-sensitive hashing functions. Bloom filters enable to speed up the search for a specified keyword using a time-space trade-off. We use locality-sensitive hashing functions to speed the search for the (appro\-ximate-)nearest neighbour of an element in a reference set. Combining these primitives enables to efficiently use cryptographic methods on biometric templates, and to achieve error-tolerant searchable encryption. 

\subsection{Organization} 

In Section \ref{sec:sec} we describe the biometric identification architecture that we consider and explain our security objectives to reach. Section \ref{sec:model} introduces the security model for the cryptographic primitives that we use, based on the new concept of Error-Tolerant Searchable Encryption. We introduce the different functions used for our proposition in Section \ref{sec:def}. We give in Section \ref{sec:constr} a step-by-step construction of an error-tolerant searchable scheme, together with its security analysis. Application to biometric identification is explained in Section \ref{sec:ident} and Section \ref{sec:pract} gives a practical illustration with IrisCodes. Section \ref{sec:ccl} concludes.

An additional property of symmetric privacy is analyzed in Appendix \ref{sec:achievingSymReqPriv}.

\section{Architecture for Biometric Identification}
\label{sec:sec}
	\subsection{Introduction to Biometric Identification} \label{sec:def:bio}

For a given biometrics technology, such as the fingerprint or the iris, let 
$B$ be the set of all possible corresponding biometric features -- i.e. data which are captured by biometric sensors. For biometric recognition, a matching algorithm $m:B\times B\rightarrow \R$ is used to compute a dissimilarity score between two data. Its goal is to differentiate similar data from different ones:
	\begin{defin}
		A biometric template $b\in B$ is the result of a measurement from someone's biometrics thanks to a sensor.  For a specific user whose biometrics is $\beta$, we note $b \leftarrow \beta$ the fact that $b$ is a measure of $\beta$.
		
		Two different measures of the same user $b, b' \leftarrow \beta$ have with high probability a \emph{small} score $m(b,b')$; measures of different users $b_1 \leftarrow \beta_1$, $b_2 \leftarrow \beta_2$ have a \emph{large} value $m(b_1,b_2)$.
	\end{defin}
	
		In practice, some thresholds $\lambda_{min}$, $\lambda_{max}$ are chosen and the score is considered as \emph{small} (resp. \emph{large}) if it is less (resp. greater) than the threshold $\lambda_{min}$ (resp. $\lambda_{max}$).  This score is usually enough to determine with some precision if two measures correspond to the same user or not. 
	Errors, called False Reject and False Acceptance, are possible but this problem is outside the scope of our paper. 	
	
		In the following, we restrict ourselves to $B = \{0,1\}^N$ equipped with the Hamming distance $d$. A biometric template $b\in B$ is the result of a measurement from someone's biometrics thanks to a sensor.	Two different measures $b,b'$ of the same user $\mathcal{U}$ are with high probability at a Hamming distance $d(b,b') \leq \lambda_{min}$ ; measures $b_1, b_2$ of different users $\mathcal{U}_1,\mathcal{U}_2$ are at a Hamming distance $d(b_1, b_2) > \lambda_{max}$. In this case, the matching algorithm $m$ simply consists in evaluating the Hamming distance.
		
				\begin{remark}
	For instance, iris biometric features are binary vectors of length $2048$ when coded as {\em IrisCodes} following \cite{Daugman93}. In this case of IrisCode \cite{Daugman93}, the matching algorithm $m$ is related to the computation of a Hamming distance between two IrisCodes. 
	\end{remark}

	 A {\em biometric identification system} -- also called a {\em one-to-many} biometric system -- recognizes a person among a collection of templates. 
	 A system is given by a reference data set $D\subset B$ and a identification function $\mathsf{id}:B\rightarrow \mathcal{P}(D)$. On input $b_{new}$, the system outputs a subset $C$ of $D$ containing biometric templates $b_{ref}\in D$ such that the matching score between $b_{new}$ and $b_{ref}$ is small. This means that $b_{new}$ and $b_{ref}$ possibly corresponds to the same person. $C$ is the $\emptyset$ if no such template can be found; the size of $C$ depends on the accuracy of the system. With pseudo-identities (either real identities of persons or pseudonyms) registered together with the reference templates in $D$, the set $C$ gives a list of candidates for the pseudo-identity of the person associated to $b_{new}$.

\subsection{Architecture}
\label{archi} 
Our general model for biometric identification relies on the following entities.
	\begin{itemize}
		\item Human users $\mathcal{U}_i$: a set of $N$ users are registered thanks to a sample of their biometrics $\beta_i$ and pseudo-identities $ID_i$, more human users $\mathcal{U}_j$ ($j >N$) represent possible impostors with biometrics $\beta_j$.
		\item Sensor client $\mathcal{SC}$: a device that extracts the biometric template from $\beta_i$.
	
\item Identity Provider $\mathcal{IP}$: replies to queries sent by $\mathcal{SC}$ by providing an identity,
		\item Database $\mathcal{DB}$: stores the biometric data.
	\end{itemize}
	
	\begin{remark}
	Here the sensor client is a client which captures the raw image of a biometric data and extracts its characteristics to output a so-called biometric template. Consequently, we assume that the sensor client is always honest and trusted by all other components. Indeed, as biometrics are public information, additional credentials are always required to establish security links in order to prevent some well-known attacks (e.g. replay attacks) and to ensure that, with a high probability, the biometric template captured by the sensor and used in the system is from a living human user. In other words, we assume that it is difficult to produce a fake biometric template that can be accepted by the sensor. 
	\end{remark}

In an identification system, we have two main services:
\begin{enumerate}
	\item $\mathsf{Enrolment}$ registers users 
	thanks to their physiological characteristics (for a user $\mathcal{U}_i$, it requires a biometric sample $b_i\leftarrow \beta_i$ and its identity $ID_i$)
	\item $\mathsf{Identification}$ answers to a request by returning a subset of the data which was registered 
\end{enumerate}

The enrolment service can be run each time a new user has to be registered. Depending on the application, the identification service can output either the identity of the candidates or their reference templates.

	As protection against outsiders, such as eavesdroppers, can be achieved with classical cryptographic techniques, our main objective is the protection of the data against insiders. In particular we assume that no attacker is able %to read communications between two parties, nor 
	to interfere with these communications.

\subsection{Informal Objectives}\label{sec:informal} 
We here formulate the properties we would like to achieve in order to meet good privacy standards.

\begin{cond}
%\label{cond:complete}
When the biometric identification system is dealing with the identification of a template $b$ coming from the registered user $\mathcal{U}_i$ with identity $ID_i$, it should return
%is \emph{complete} if, 
a subset containing a reference to $(ID_i,b_i)$ except for a negligible probability.
\end{cond}

\begin{cond}
	%\label{cond:sound}
When the system is dealing with the identification of a  template $b$ coming from an unregistered user, it should return the empty set $\emptyset$ except for a negligible probability.
\end{cond}

We do not want a malicious database to be able to link an identity to a biometric template, nor to be able to make relations between different identities. 

\begin{cond}
	%\label{cond:idPriv}
	The database  $\mathcal{DB}$ should not be able to distinguish two enrolled biometric data.
\end{cond}

Another desired property is the fact that the database knows nothing of the identity of the user who goes through the identification process, for example, to avoid unwanted statistics.

\begin{cond}
	%\label{cond:transAnon}
		The database  $\mathcal{DB}$ should not be able to guess which identification request is executed.
	\end{cond}

\section{Security Model for Error-Tolerant Searchable Encryption}
	\label{sec:model}

In this section, we describe a formal model for an error-tolerant searchable encryption protocol. A specific construction fitting in this model is described in Section \ref{sec:constr}. This scheme enables to approximately search and retrieve a message stored in a database, i.e. with some error-tolerance on the request. This is in fact a problem quite close to biometric identification and the corresponding cryptographic primitives are thus used in our system, cf. Section \ref{sec:ident}.

In the sequel, we note $\discretAB m n$ the set of all integers between $m$ and $n$ (inclusive).

\subsection{Entities for the Protocol}

Our primitive models the interactions between users that store and retrieve information, and a remote server. We distinguish the user who stores the data from the one who wants to get it. This leads to three entities:
\begin{itemize}
\item The server $\Se$: a remote storage system. As the server is untrusted, 
%-- it can be an outsourcing server, a free service offered by some mail provider --  
we consider the content to be public. Communications to and from this server are also subject to eavesdropping,
\item The sender $\X$ incrementally creates the database, by sending data to $\Se$,
\item The receiver $\Y$ makes queries to the server $\Se$.
\end{itemize}

In a latter part (Sec. \ref{sec:ident}), we integrate our cryptographic protocols into our biometric identification system. This doing, we merge the entities defined in Sec. \ref{archi} and those just previously introduced.

We emphasize that $\X$ and $\Y$ are not necessarily the same user, as $\X$ has full knowledge of the database he created whereas $\Y$ knows only what he receives from $\Se$.

\subsection{Definition of the Primitives} \label{sec:primitiv} \label{sec:prim}

In the sequel, messages are binary strings of a fixed length $N$, and $d(x_1,x_2)$ the Hamming Distance between $x_1,x_2 \in \{0,1\}^N$ is the canonical distance, \textit{i.e.} the number of positions in $\discretAB{1}{N}$ in which $x_1$ and $x_2$ differ.

Here comes a formal definition of the primitives that enable to perform an error-tolerant searchable encryption; this definition cannot be parted from the definition of Completeness($\lambda_{min}$) and $\epsilon$-Soundness($\lambda_{max}$), which follows.

\begin{defin}
	A \emph{$(\epsilon, \lambda_{{min}},\lambda_{{max}})$-Public Key Error-Tolerant Searchable Encryption} is obtained with the following probabilistic polynomial-time methods:
	\begin{itemize}
		\item $\mathsf{KeyGen}(1^k)$ initializes the system, and outputs public and private keys $(pk,sk)$; $k$ is the security parameter. The public key $pk$ is used to store data on a server, and the secret key $sk$ is used to retrieve information from that server.
		\item $\mathsf{Send}_{\X,\Se}(x, pk)$ is a protocol in which $\X$ sends to $\Se$ the data $x\in\{0, 1\}^N$ to be stored on the storage system. At the end of the protocol, $\Se$ associated an identifier to $x$, noted $\varphi(x)$.
		%At the end of this protocol, $\X$ detains a tag associated to $x$, noted $\varphi(x)$.
		\item $\mathsf{Retrieve}_{\Y,\Se}(x',sk)$ is a protocol in which, given a fresh data $x'\in\{0, 1\}^N$, $\Y$ asks for the identifiers of all data that are stored on $\Se$ and are close to $x'$, with Completeness($\lambda_{min}$) and $\epsilon$-Soundness($\lambda_{max}$). This outputs a set of identifiers, noted $\Phi(x')$. %TODO : peut être reformuler ça.
	\end{itemize}
\end{defin}

These definitions are comforted by the condition \ref{cond:complete} of Section \ref{sec:secreq} that defines Completeness and $\epsilon$-Soundness for the parameters already introduced in Section \ref{sec:def:bio}, $\lambda_{min}, \lambda_{max}$. In a few words, Completeness implies that a registered message $x$ is indeed found if the query word $x'$ is at a distance less than $\lambda_{min}$ from $x$, while $\epsilon$-Soundness means that with probability greater than $1-\epsilon$, no message at a distance greater than $\lambda_{max}$ from $x'$ will be returned.

The $\mathsf{Send}$ protocol produces an output $\varphi(x)$ that identifies the data $x$. This output $\varphi(x)$ is meant to be a unique identifier, which is a binary string of undetermined length -- in other words, elements of $\{0, 1\}^{\star}$ -- that enables to retrieve $x$. It can be a timestamp, a name or nickname, \textit{etc.} depending on the application.

\subsection{Security Requirements} \label{sec:secreq}
 
 First of all, it is important that the scheme actually works, \textit{i.e.} that the retrieval of a message near a registered one gives the correct result. This can be formalized into the following condition:

\begin{cond}[\textbf{Completeness($\lambda_{min}$), $\epsilon$-Soundness($\lambda_{max}$)}]  \label{cond:complete} \label{cond:sound}
Let $x_1, \ldots, x_p\in B=\{0, 1\}^N$ be $p$ different binary vectors, and let $x'\in B$ be another binary vector. Suppose that the system was initialized, that all the messages $x_i$ have been sent by user $\X$ to the system $\Se$ with identifiers $\varphi(x_i)$, and that user $\Y$ retrieved the set of identifiers $\Phi(x')$ associated to $x'$.

\begin{enumerate}
	\item The scheme is said to be \textbf{complete} if the identifiers of all the $x_i$ that are near $x'$ are almost all in the resulting set $\Phi(x')$, \textit{i.e.} if $$\eta_c = \Ppp{x'}{\exists i \text{ s.t. } d(x', x_i)\leq \lambda_{min} \text{ and } \varphi(x_i) \notin \Phi(x')}$$ is negligible.
	\item The scheme is said to be \textbf{$\epsilon$-sound} if the probability of finding an unwanted result in $\Phi(x')$, i.e. 
	\begin{eqnarray*}
	\eta_s&=&\Pr_{x'}\left[\exists i \in \discretAB{1}{p} \text{ s.t. } d(x', x_i)> \lambda_{max} \right.
	\left. \text{ and } \varphi(x_i) \in \Phi(x')\right],
	\end{eqnarray*} is bounded by $\epsilon$.
\end{enumerate}
\end{cond}

The first condition simply means that registered data is effectively retrieved if the input is close. $\eta_c$ expresses the probability of failure of this $\mathsf{Retrieve}$ operation. 

The second condition means that only the close messages are retrieved, thus limiting false alarms. $\eta_s$ measures the reliability of the $\mathsf{Retrieve}$ query, \textit{i.e.} if all the results are identifiers of messages near to $x'$.

These two properties (\emph{Completeness} and $\epsilon$-\emph{Soundness}) are sufficient to have a working set of primitives which allows to make approximate queries on a remote storage server. The following conditions, namely \emph{Sender Privacy} and \emph{Receiver Privacy}, ensure that the data stored in the server is secure, and that communications can be done on an untrusted network. 

\begin{cond}[\textbf{Sender Privacy}] \label{cond:messPriv}\label{cond:senderPriv}
The scheme is said to respect \emph{Sender Privacy} if the advantage of any malicious server is negligible in the $\mathsf{Exp}_{\mathcal{A}}^{\text{Sender Privacy}}$ experiment, described below. Here, $\mathcal{A}$ is an ``honest-but-curious'' opponent taking the place of $\Se$, and $\mathcal{C}$ is a challenger at the user side.
	\begin{small}
	$$
		\begin{array}{ll}
				\mathsf{Exp}_{\mathcal{A}}^{\text{Sender Privacy}} & \\
			& \hspace{-2cm}%\hspace{-2.7cm}
				\vline 
				\begin{array}{clclc}
					1.	& (pk,sk)	& \lar	& \mathsf{KeyGen}(1^k)			& (\mathcal{C})	\\
					2.	& \{x_2, \ldots, x_\Omega\}	& \lar	& \mathcal{A}	 &  (\mathcal{A})\\
					3.	& \varphi(x_i) &\lar  & \mathsf{Send}_{\X, \Se}(x_i, pk)	& (\mathcal{C})	\\
					4.	& \{x_0, x_1\}	& \lar	& \mathcal{A}	& (\mathcal{A})\\
					%5.	& \varphi(x_e) & \lar & \mathsf{Send}_{\X, \Se}(x_e, pk),  e\in_R\{0, 1\}	& (\mathcal{C})	\\
					5.	& \varphi(x_e) & \lar & \mathsf{Send}_{\X, \Se}(x_e, pk),  	& (\mathcal{C})	\\
							&							 &			& e\in_R\{0, 1\} \\
					6.	& \text{Repeat steps } (2,3)				& 			&					\\
					7.	& e'\in \{0, 1\}								& \lar	& \mathcal{A} & (\mathcal{A})
				\end{array} 
		\end{array}
	$$
	\end{small}
The advantage of the adversary is $|\Pp{e'=e} - \frac{1}{2}|$.
\end{cond}

Informally, in a first step, the adversary receives $\mathsf{Send}$ requests that he chose himself; $\mathcal{A}$ then looks for a couple $(x_0,x_1)$ of messages on which he should have an advantage. $\mathcal{C}$ chooses one of the two messages, and the adversary must guess, by receiving the $\mathsf{Send}$ requests, which one of $x_0$ or $x_1$ it was.

This condition permits to have privacy on the content stored on the server. The content that the sender transmits is protected, justifying the title ``Sender Privacy''. %In a biometric application, this denomination is justified twice, as the data that is sent directly concerns the privacy of the sender.

Another important privacy aspect is the secrecy of the data that is retrieved. We do not want the server to have information on the fresh data $x'$ that is queried; this is expressed by the following condition.

\begin{cond}[\textbf{Receiver Privacy}] \label{cond:reqPriv} \label{cond:receiverPriv}
The scheme is said to respect \emph{Receiver Privacy} if the advantage of any malicious server is negligible in the $\mathsf{Exp}_{\mathcal{A}}^{\text{Receiver Privacy}}$ experiment described below. As in the previous condition, $\mathcal{A}$ denotes the ``honest-but-curious'' opponent taking the place of $\Se$, and $\mathcal{C}$ the challenger at the user side.
	\begin{small}
	$$
		\begin{array}{ll}
				\mathsf{Exp}_{\mathcal{A}}^{\text{Receiver Privacy}} & \\
			& \hspace{-2cm}\vline 
				\begin{array}{clclc}
					1.	& (pk,sk) & \lar & \mathsf{KeyGen}(1^k)			& (\mathcal{C})	\\
					2.	& \{x_1, \ldots, x_\Omega\} & \lar	& \mathcal{A}	 & (\mathcal{A}) \\
						& \multicolumn{3}{l}{d(x_i, x_j)> \lambda_{max}, \forall i,j\in\discretAB{1}{\Omega}}	 & \\
					3.	& \varphi(x_i), (i\in\discretAB{1}{\Omega}) 			& \lar & \mathsf{Send}_{\X, \Se}(x_i, pk)	& (\mathcal{C})	\\
					4.	& \{x'_2, \ldots, x'_p\}					& \lar	& \mathcal{A}	 & (\mathcal{A})\\
					5.	& \Phi(x'_j),  (j\in\discretAB{2}{p}) & \lar & \mathsf{Retrieve}_{\Y, \Se}(x'_j, sk)	&	(\mathcal{C})	\\
					6.	& (x'_0, x'_1)	& \lar	& \mathcal{A}	& (\mathcal{A})\\
					7.	& \Phi(x'_e) & \lar & \mathsf{Retrieve}_{\Y, \Se}(x'_e, sk), &(\mathcal{C})\\
							&						 &			& e\in_R\{0,1\} 	\\
					8.	& \text{Repeat steps } (4,5)				& 			&					\\
					9.	& e'\in \{0, 1\}								& \lar	& \mathcal{A} & (\mathcal{A})
				\end{array}
		\end{array}
	$$
	\end{small}
The advantage of the adversary is $|\Pp{e'=e} - \frac{1}{2}|$.
\end{cond}
This condition is the mirror image of the previous one. It transposes the idea that the receiver $\Y$ can make his queries to $\Se$ without leaking information on their content. The processing of the experiment is the same as the Sender Privacy experiment, except that $\mathcal{A}$ has to distinguish between $\mathsf{Retrieve}$ queries instead of $\mathsf{Send}$ queries.

\begin{remark}
Conditions \ref{cond:senderPriv} and \ref{cond:receiverPriv} are the transposition of their homonym statement in \cite{BonehKOS07}. They aim for the same goal, \textit{i.e.} privacy -- against the server -- of the data that is registered first, then looked for. 
\end{remark}

Section \ref{sec:constr} is dedicated to give a construction that fits these security conditions.

\section{Our Data Structure for Approximate Searching}
	\label{sec:def}
	
		After the recall of the notions of locality-sensitive hashing and Bloom filters, we introduce a new structure which enables approximate searching by combining both notions. We end this section with the introduction of some classical cryptographic protocols.
	
	In the sequel, we denote $[a,b]$ the interval of all real values between $a$ and $b$ (inclusive).

	\subsection{Locality-Sensitive Hashing} \label{sec:def:lsh}	

We first consider the following problem:
	
	\begin{prob}[Approximate Nearest Neighbour Problem]
		Given a set $P$ of points in the metric space $(B, d)$ pre-process $P$ to efficiently answer queries. The answer of a query $x$ is a point $p_x\in P$ such that $d(x, p_x) \leq (1+\epsilon) \min_{p\in P} d(x,p)$.
	\end{prob}

	This problem has been widely studied over the last decades; reviews on the subject include \cite{indyk04nearest}. However, most algorithms proposed to solve the matter consider real spaces over the $l_p$ distance, which is not relevant in our case. 	A way to search the approximate nearest neighbour in a Hamming space is to use a generic construction called locality-sensitive hashing. It looks for hash functions (not cryptographic ones) that give the same result for near points, as defined in \cite{IndykM98}:
	\begin{defin}[\cite{IndykM98}]
		Let $B$ be a metric space, $U$ a set with a smaller dimensionality, $r_1, r_2 \in \mathbb{R}$ with $r_1<r_2$, $p_1, p_2\in [0,1]$ with $p_1> p_2$. A family $H=\{h_1, \ldots, h_\mu\}, h_i: B \rightarrow U$, is \emph{$(r_1, r_2, p_1, p_2)$-LSH (Locality-Sensitive Hashing)}, if for all $h\in H, x,x'\in B$, $\Pp{h(x)=h(x')} >p_1$ (resp. $\Pp{h(x)=h(x')} <p_2$) if $d_B(x,x')<r_1$ (resp. $d_B(x,x')>r_2$).
	\end{defin}
	
	Such functions reduce the differences occurring between similar data with high probability, whereas distant data should remain significantly remote. 
	
		A noticeable example of a LSH family was proposed by Kushilevitz~\emph{et al.} in \cite{KushilevitzOR98}; see also \cite{KirschMitzen06,IndykM98,AndoniI08}.
	
	\subsection{Bloom Filters}
	\label{sec:bloom}
	
	As introduced by Bloom in \cite{Bloom70}, a set of Bloom filters is a data structure used for answering set membership queries. 
	
	\begin{defin}
		\label{def:bloom}
		Let $D$ be a finite subset of $Y$. For a collection of $\nu$ (independent) hash functions $H'=\{h'_1, \ldots, h'_\nu \}$, with each $h'_i: Y \rightarrow \discretAB 1 m$ , the induced \emph{$(\nu,m)$-Bloom filter} is $H$, together with an array $(t_1, \ldots, t_m)\in \{0,1\}^m$, defined as:
		$$t_\alpha = \left \{ \begin{aligned}
														1 & \text{ if } \exists i \in \discretAB 1 \nu , y\in D \text{ s.t. } h'_i(y)=\alpha\\
														0 & \text{ otherwise}
													\end{aligned} \right .$$
	\end{defin}
	
	With this setting, testing if $y$ is in $D$ is the same as checking if for all $i \in \discretAB 1 \nu,  t_{h'_i(y)}=1$. The best setting for the filter is that the involved hash function be as randomized as possible,  
	in order to fill all the buckets $t_\alpha$. %However, cryptographic hash functions are usually ignored for Bloom filters, as the aim of the filter is to speed up the search for a given bit-string. 
	
	In this setting, some false positive may happen, \textit{i.e.} it is possible for all $t_{h'_i(y)}$ to be set to $1$ and $y \notin D$. This event is well known, and the probability for a query to be a false positive is: $\left ( 1- \left ( 1- \frac{\nu}{m} \right )^{|D|} \right ) ^\nu$.
	
	This probability can be made as small as needed. 
	On the other hand, no false negative is enabled.
	
We work here with the \emph{Bloom filters with storage} (\highlight{BFS}) defined in \cite{BonehKOS07} as an extension of Bloom filters. Their aim is to give not only the result of the set membership test, but also an index associated to the element. The iterative definition below introduces these objects and the notion of \emph{tags} and \emph{buckets} which are used in the construction. 
	
	\begin{defin}[Bloom Filter with Storage, \cite{BonehKOS07}]\label{def:bfs}
	Let $D$ be a finite subset of a set $Y$. For a collection of $\nu$ hash functions $H'=\{h'_1, \ldots, h'_\nu \}$, with each $h'_j: Y \rightarrow \discretAB 1 m$, a set $V$ of \emph{tags} associated to $D$ with a \emph{tagging} function $\psi:D\rightarrow \mathcal{P}(V)$. 
		A \emph{$(\nu,m)$-Bloom Filter with Storage } is $H'$, together with an array of subsets $(T_1, \ldots, T_m)$ of $V$, called \emph{buckets}, iteratively defined as: 
		\begin{enumerate}
			\item $\forall i\in \discretAB{1}{m}, T_i \leftarrow \emptyset$,
			\item $\forall y\in D, \forall j\in \discretAB{1}{\nu}, $ update the bucket $T_\alpha$ with $T_\alpha \leftarrow T_\alpha \cup \psi(y)$ where $\alpha = h'_j(y)$.
		\end{enumerate}
	\end{defin}

\begin{figure}[hbtp!]
\centering
{\scriptsize
\psfrag{ca}{$\cdots$}\psfrag{cb}{$\cdots$}
\psfrag{c2}{\hspace{-0.25cm}$\psi(y_1)$}\psfrag{c3}{\hspace{-0.25cm}$\psi(y_1)$}\psfrag{c4}{\hspace{-0.25cm}$\psi(y_1)$}
\psfrag{c1}{\hspace{-0.25cm}$\psi(y_2)$}\psfrag{c22}{\hspace{-0.25cm}$\psi(y_2)$}\psfrag{c32}{\hspace{-0.25cm}$\psi(y_2)$}
\psfrag{c23}{\hspace{-0.25cm}$\psi(y_3)$}\psfrag{c42}{\hspace{-0.25cm}$\psi(y_3)$}\psfrag{c5}{\hspace{-0.25cm}$\psi(y_3)$}
\psfrag{i2}{\hspace{-0.35cm}$h'_2(y_3)=2$}\psfrag{i4}{\hspace{-0.35cm}$h'_1(y_3)=\alpha$}\psfrag{i5}{\hspace{-0.35cm}$h'_3(y_3)=m$}
\psfrag{empty}{$\emptyset$}
\includegraphics[width=\columnwidth]{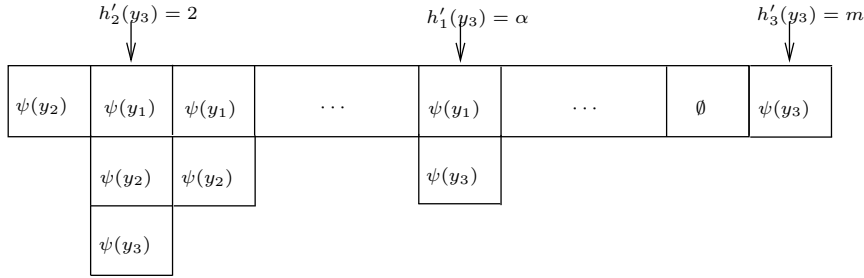}}
\caption{Construction of Bloom Filters with Storage}		\label{fig:bfs}
\end{figure}

In other words, the bucket structure is empty at first, and for each element $y\in D$ to be indexed, we add to the bucket $T_\alpha$ all the tags associated to $y$.	
	 Construction of such a structure is illustrated in Fig. \ref{fig:bfs}.
	 	
\begin{ex}
	 In Fig. \ref{fig:bfs}, assume that $D=\{y_1,y_2,y_3\}$ and $\nu=3$, the tags associated to $y_1$ (resp. $y_2$) have already been incorporated into the buckets $T_2$, $T_3$ and $T_\alpha$ (resp. $T_1$, $T_2$ and $T_3$) so that $T_1=\{\psi(y_2)\}$, $T_2=T_3= \{\psi(y_1),\psi(y_2)\}$, $T_\alpha=\{\psi(y_1)\}$ and $T_i=\emptyset$ otherwise. We are now treating the case of $y_3$:
	 \begin{itemize}
	 \item $h'_1(y_3)=\alpha$ so $T_\alpha \leftarrow T_\alpha \cup \{\psi(y_3)\}$, i.e. $T_\alpha=\{\psi(y_1),\psi(y_3)\}$;
	 \item $h'_2(y_3)=2$ so $T_2 \leftarrow T_2 \cup \{\psi(y_3)\}$, i.e. $T_2=\{\psi(y_1),\psi(y_2),\psi(y_3)\}$;
	 \item $h'_3(y_3)=m$ so $T_m \leftarrow T_m \cup \{\psi(y_3)\}$, i.e. $T_m=\{\psi(y_3)\}$.
	 \end{itemize}
\end{ex}

	This construction enables to retrieve a set of tags associated to an element $y\in D$: it is designed to obtain $\psi(y)$, the set of tags associated to $y$, by computing  $\bigcap_{j=1}^{\nu} T_{h'_j(y)}$.
	For instance, in the previous example, $\bigcap_{j=1}^{\nu} T_{h'_j(y_3)}=T_2 \cap T_\alpha \cap T_m=\{\psi(y_3)\}$. %Fig. \ref{fig:bfs}
This intersection may capture inappropriate tags, but the choice of relevant hash functions and increasing their number allow to reduce the probability of that event. These properties are summed up in the following lemma.

	\begin{lemma}[\cite{Bloom70}] \label{lemma:bloom}
	Let $(H', T_1, \ldots, T_m)$ be a $(\nu, m)$-Bloom filter with storage indexing a set $D$ with tags from a tag set $V$. Then, for $y\in D$, the following properties hold:
		\begin{itemize}
			\item $\psi(y) \subset T(y) = \bigcap_{j=1}^{\nu} T_{h'_j(y)}$, \textit{i.e.} each of $y$'s tag is retrieved,
			\item the probability for a false positive $t\in V$ is $\Pp{t\in T(y) \text{ and } t\not\in \psi(y)} = \left ( 1- \left ( 1- \frac{\nu}{m} \right )^{|D|} \right ) ^\nu$.
		\end{itemize}
	\end{lemma}

	\subsection{Combining BFS and LSH}
	\label{sec:Blsh}
	We want to apply Bloom filters to data that are very likely to vary. To this aim, we first apply LSH-families as input to Bloom filters. 
	
	We choose $\mu$ hash functions from an adequate LSH family $h_1, \ldots, h_\mu:B \rightarrow \{0, 1\}^t$, and $\nu$ hash functions dedicated to a Bloom filter with Storage $h'_1, \ldots, h'_\nu: \{0, 1\}^t \times \discretAB 1 \mu \rightarrow \discretAB 1 m$. The LSH family is denoted $H$, and $H'$ is the BFS one.
	To obtain a BFS with locality-sensitive functionality, we use composite $\mu \times \nu$ hash functions induced by both families.

	We define $h^c_{(j,i)}:B\rightarrow \discretAB 1 m$ the corresponding composite functions ($c$ stands for composite) with  $h^c_{(j,i)}(y)=h'_j(h_i(y)\parallel i)$. Let $H^c=\{h^c_{(j,i)}, (j,i)\in \discretAB{1}{\nu} \times \discretAB{1}{\mu}\}$ the set of all these functions. 

	To sum up, we modify the update of the buckets in Def. \ref{def:bfs} by $\alpha = h'_j(h_i(y)\parallel i)$. Later on, to recover tags related to an approximate query $x'\in B$, all we have to consider is $\bigcap_{j=1}^{\nu}\bigcap_{i=1}^{\mu} T_{h'_j(h_i(x')\parallel i)}$. Indeed, if $x$ and $x'$ are close enough, then the LSH functions give the same results on $x$ and $x'$, effectively providing a Bloom filter with storage that has the LSH property. This property is numerically estimated in the following lemma:
	
	\begin{lemma} \label{lemma:BLSH}
		Let $H, H', H^c$ be families constructed in this setting. Let $x, x'\in B$ be two binary vectors. Assume that $H$ is $(\lambda_{min}, \lambda_{max}, \epsilon_1, \epsilon_2)$-LSH from $B$ to $\{0, 1\}^t$; assume that $H'$ is a family of $\nu$ pseudo-random hash functions. If the tagging function $\psi$ associates only one tag per element, then the following properties stand:
		\begin{enumerate}
			\item If $x$ and $x'$ are far enough, then except with a small probability, $\psi(x')$ does not intersect all the buckets that index $x$, \textit{i.e.} 
			\begin{small}
			\begin{eqnarray*} \Ppp{x'}{\psi(x') \subset \bigcap_{h^c\in H^c}  T_{h^c(x)} \text{ and } d(x, x')  \geq \lambda_{max}}  \leq \left(\epsilon_2 + (1-\epsilon_2)\frac{1}{m}\right)^{|H^c|},
			\end{eqnarray*}
			\end{small}
			\item If $x$ and $x'$ are close enough, then except with a small probability, $\psi(x')$ is in all the buckets that index $x'$, \textit{i.e.} 
			\begin{small}\begin{eqnarray*}\Ppp{x'}{\psi(x')  \not\subset \bigcap_{h^c\in H^c}  T_{h^c(x)} \text{ and } d(x, x') \leq \lambda_{min}}  \leq  1 - \left(1-\epsilon_1 \right) ^{|H^c|}.
			\end{eqnarray*}
			\end{small}
		\end{enumerate}
	\end{lemma}
	
	Note that this lemma used the simplified hypothesis that $\forall x, |\psi(x)|=1$, which means that there is only one tag per vector. This has a direct application in Section \ref{sec:secProp}. In practice, $\psi(x)$ can be a unique handle for $x$.%and is not prohibitive in practice.
	
	{\em Sketch of proof.}\,
	The first part of the lemma expresses the fact that if $d(x,x')\geq \lambda_{max}$, due to the composition of a LSH function with a pseudorandom function, the collision probability is $\frac{1}{m}$. Indeed, if $h'_1(y_1) = h'_2(y_2)$, either $y_1=y_2$ and $h'_1 = h'_2$, or there is a collision of two independent pseudo-random hash function. In our case, if $y_1=y_2$, that means that $y_1 = h_{i_1}(x)||i_1$ and $y_2 = h_{i_2}(x')||i_2$. To these vectors to be the same, $i_1=i_2$ and $h_{i_1}(x)=h_{i_2}(x')$, which happens with probability $\epsilon_2$.
	
	The second part of the lemma says that for each $h^c\in H^c$, $h^c(x)$ and $h^c(x')$ are the same with probability $1-\epsilon_1$. Combining the incremental construction of the $T_i$ with this property gives the lemma.
	\findemo
	
	\subsection{Cryptographic Primitives} \label{sec:mod:prim}

\paragraph{Public Key Cryptosystem}
	Our construction requires a semantically secure public key cryptosystem -- as defined in \cite{GoldwasserM84}, see for instance \cite{Gamal84,Pa99} -- to store some encrypted data in the database. Encryption function is noted $\Enc$ and decryption function $\Dec$, the use of the keys is implicit. 
	An encryption scheme is said to be semantically secure (against a chosen plaintext attack, also noted IND-CPA \cite{GoldwasserM84}) if an adversary without access to the secret key $sk$, cannot distinguish between the encryptions of a message $x_0$ and a message $x_1$.

\paragraph{Private Information Retrieval Protocols} \label{sec:pir}

	A primitive that enables privacy-ensuring queries to databases is Private Information Retrieval protocol (PIR, \cite{ChorKGS98}). Its goal is to retrieve a specific information from a remote server in such a way that he does not know which data was sent. This is done through a method $\mathsf{Query}^{PIR}_{\Y, \Se}(a)$, that allows $\Y$ to recover the element stored at index $a$ in $\Se$ by running the PIR protocol.

	Suppose a database is constituted with $M$ bits  $X = x_1, ...,x_M$. To be secure, the protocol should satisfy the following properties~\cite{GertnerIKM98}:
\begin{itemize}
\item {\bf Soundness:} When the user and the database follow the protocol, the result of the request is exactly the requested bit.
\item {\bf User Privacy:} For all $X \in \{0,1\}^M$, for $1\leq i,j \leq M$, for any algorithm used by the database,  it cannot distinguish with a non-negligible probability the difference between the requests of index $i$ and $j$.
\end{itemize}
	
Among the known constructions of computational secure PIR, block-based PIR -- i.e. working on block of bits -- allows to efficiently reduce the cost.  The best performances are from Gentry and Ramzan~\cite{GentryR05} and   Lipmaa~\cite{Lipmaa05} with a communication complexity polynomial in the logarithm of $M$. Surveys of the subject are available in~\cite{gasarch:1,OstrovskyS97}.

	Some PIR protocols are called Symmetric Private Information Retrieval, when they comply with the \textbf{Data Privacy} requirement \cite{GertnerIKM98}. This condition states that the querier cannot distinguish between a database that possesses only the information he requested, and a regular one; in other words, that the querier do not get more information that what he asked.

\paragraph{Private Information Storage (PIS) Protocols} \label{rem:pis}
	PIR protocols enable to retrieve information of a database. A Private Information Storage (PIS) protocol \cite{OstrovskyS97} is a protocol that enables to write information in a database with properties that are similar to that of PIR. The goal is to prevent the database from knowing the content of the information that is being stored; for detailed description of such protocols, see \cite{BonehKOS07,OstSke07a}.
	
	Such a protocol provides a method $\mathsf{update}(val,index)$, which takes as input an element and a database index, and puts the value $val$ into the database entry $index$. To be secure, the protocol must also satisfy the Soundness and User Privacy properties, meaning that 1. $\mathsf{update_{BF}}$ does update the database with the appropriate value, and 2. any algorithm run by the database cannot distinguish between the writing requests of $(val_i, ind_i)$ and $(val_j, ind_j)$.

\section{Our Construction for Error-Tolerant Searchable Encryption} \label{sec:constr}

\subsection{Technical Description}
 \label{sec:detail}

Our searching scheme uses all the tools we described in the previous section. As we will see in section \ref{sec:secProp}, this enables to meet the privacy requirements of section \ref{sec:secreq}. More precisely:
\begin{itemize}
\item We pick a family $H'$ of functions: $h':\{0,1\}^t\times \discretAB{1}{|H|} \rightarrow \discretAB{1}{m}$, adapted to a Bloom filter structure,
\item We choose a family $H$ of functions: $h:\{0,1\}^N \rightarrow \{0,1\}^t$ that have the LSH property,
\item From these two families, we deduce a family $H^c$ of functions $h^c:\{0,1\}^N \rightarrow \discretAB{1}{m}$ as specified in Sec. \ref{sec:Blsh},
\item We use a semantically secure public key cryptosystem $\left ( \mathsf{Setup}, \Enc, \Dec \right )$ \cite{GoldwasserM84},
\item We use a PIR protocol with query function $\mathsf{Query}^{PIR}_{\Y, \Se}$.
\item We use a PIS function $\mathsf{update_{BF}}(val,i)$ that adds $val$ to the $i$-th bucket of the Bloom filter, see Sec. \ref{rem:pis}.%. For privacy purposes, we use a method that does not leak information on its arguments, \textit{e.g.} to meet the description of a Private Information Storage function \cite{OstrovskyS97}, see Remark \ref{rem:pis}.
\end{itemize}

Here come the details of the implementation. In a few words, storage and indexing of the data are separated, so that it becomes feasible to search over the encrypted documents. Indexing is made thanks to Bloom Filters, with an extra precaution of encrypting the content of all the buckets. Finally, using our locality-sensitive hashing functions permits error-tolerance.

\subsubsection{System setup} The method $\mathsf{KeyGen}(1^k)$ initializes $m$ different buckets to $\emptyset$. The public and secret keys of the cryptosystem $(pk, sk)$ are generated by $\mathsf{Setup}(1^k)$, and $sk$ is given to $\Y$.

\subsubsection{Sending a message} 
\begin{figure}[hbtp!]
\begin{center}
\begin{scriptsize}
\psfrag{S}{$\mathcal{X}$}
\psfrag{DB}{$\mathcal{S}$}
\psfrag{s1}{1. }
\psfrag{s2}{2. }
\psfrag{m1}{\hspace{1cm} mem alloc}
\psfrag{m2}{\hspace{1cm} $\varphi(x) = \&x$}
\psfrag{m3}{Store $\Enc(x)$ at $\varphi(x)$}
\psfrag{m4}{\hspace{-0.8cm}Store $\Enc(\varphi(x))$ at all $\left\{h_i^c(x)\right\}_i$}
\includegraphics[width=0.9\linewidth]{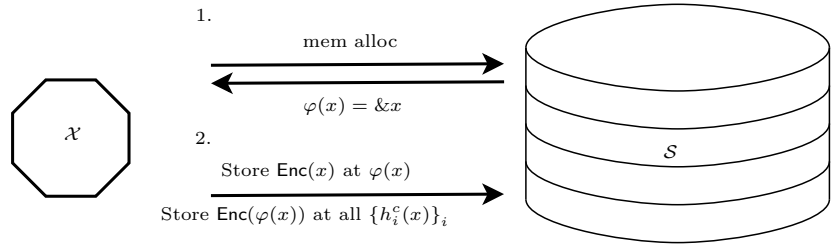}
\end{scriptsize}
\caption{Sending a message in a nutshell}
		\label{fig:enrol}
\end{center}
\end{figure}

The protocol $\mathsf{Send}_{\X, \Se}(x, pk)$ goes through the following steps (cf. Fig. \ref{fig:enrol}):
\begin{enumerate}
	%\item {\bf Identifier agreement} $\X$ and $\Se$ agree on a unique identifier $\varphi(x)$.
	\item {\bf Identifier establishment} $\Se$ attributes to $x$ a unique identifier $\varphi(x)$, and sends it to $\X$.
	\item {\bf Data storage} $\X$ sends $\Enc(x)$ to $\Se$, who stores it in a memory cell that depends on $\varphi(x)$.
	\item {\bf Data indexing} 
	\begin{itemize}
	\item $\X$ computes $h^c(x)$ for all $h^c\in H^c$, 
	\item and executes $\mathsf{update_{BF}}(\Enc(\varphi(x)), h^c(x))$ to send $\Enc(\varphi(x))$ to be added to the filter's bucket of index $h^c(x)$ on the server side.
	\end{itemize}
\end{enumerate}
Note that for privacy concerns, we complete the buckets with random data in order to get the same bucket size $l$ for the whole data structure.

The first phase (identifier establishment) is done to create an identifier that can be used to register and then retrieve $x$ from the database. For example, $\varphi(x)$ can be the time at which $\Se$ received $x$, or the first memory address that is free for the storage of $\Enc(x)$.

The third phase applies the combination of BFS and LSH functions (see Sec. \ref{sec:Blsh}) to $x$ so that it is possible to retrieve $x$ with some approximate data. This is done with the procedure described hereafter.

\subsubsection{Retrieving data} 
\begin{figure}[hbtp!]
\begin{center}
\begin{scriptsize}
\psfrag{S}{$\mathcal{Y}$}
\psfrag{DB}{$\mathcal{S}$}
\psfrag{s1}{PIR}
\psfrag{m1}{\hspace{0.5cm} $\hspace{-0.3cm}\alpha_i = h_i^c(x')$}
\psfrag{m2}{\hspace{0.7cm} $T_{\alpha_i}$}
\psfrag{m3}{\hspace{-1cm}$\bigcap_i T_{\alpha_i} = \{\varphi(x_{i_1}), \ldots, \varphi(x_{i_\gamma})\}$}
\includegraphics[width=0.9\linewidth]{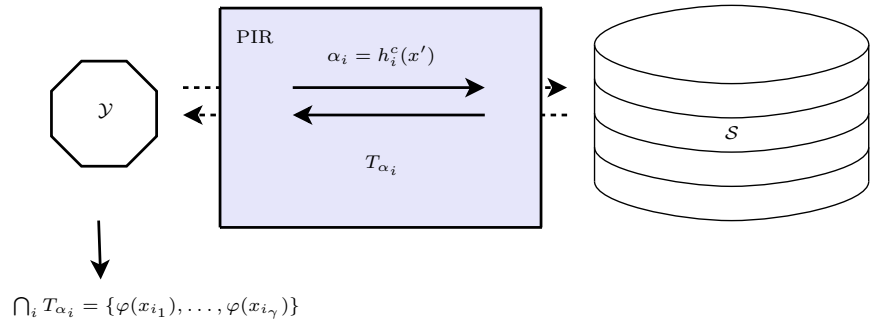}
\end{scriptsize}
\caption{Retrieving data in a nutshell}
		\label{fig:ident}
\end{center}
\end{figure}

The protocol $\mathsf{Retrieve}_{\Y, \Se}(x', sk)$ goes through the following steps (cf. Fig. \ref{fig:ident}):
\begin{enumerate}
	\item $\Y$ computes each $\alpha_i = h_i^c(x')$ for each $h_i^c \in H^c$, then executes $\mathsf{Query}^{PIR}_{\Y, \Se}(\alpha_i)$ to receive the filter bucket $T_{\alpha_i}$,
	\item $\Y$ decrypts the content of each bucket $T_{\alpha_i}$ and computes the intersection of all the $\mathsf{Dec}(T_{\alpha_i})$,
	\item This intersection is a set of identifiers $\{\varphi(x_{i_1}), \ldots, \varphi(x_{i_\gamma})\}$, which is the result of the execution of $\mathsf{Retrieve}$.
\end{enumerate}

As we can see, the retrieving process follows that of Sec. \ref{sec:Blsh}, with the noticeable differences that 1. the identifiers are always encrypted in the database, and 2. the query is made following a PIR protocol. This permits to benefit from both the Bloom filter structure, the locality-sensitive hashing, and the privacy-preserving protocols. 

The secure protocols involved do not leak information on the requests made, and the next section discusses more precisely the security properties achieved.

\subsection{Security Properties} \label{sec:secProp}

We now demonstrate that this construction faithfully achieves the security requirements we defined in Sec. \ref{sec:secreq}.

\begin{prop}[\textbf{Completeness}] \label{prop:complete}
Provided that $H$ is a $(\lambda_{min}, \lambda_{max}, \epsilon_1, \epsilon_2)$-LSH family, for a negligible $\epsilon_1$, this scheme is complete.
\end{prop}

\begin{prop}[\textbf{$\epsilon$-Soundness}] \label{prop:epsilonSound}
	Provided that $H$ is a $(\lambda_{min}, \lambda_{max}, \epsilon_1, \epsilon_2)$-LSH family from $\{0, 1\}^N$ to $\{0, 1\}^t$, and provided that the Bloom filter functions $H'$ behave like pseudo-random functions from $\{0, 1\}^t \times \discretAB{1}{|H|}$ to $\discretAB{1}{m}$, then the scheme is $\epsilon$-sound, with:
	\begin{small}$$\epsilon = \left(\epsilon_2 + (1-\epsilon_2)\frac{1}{m}\right)^{|H^c|}$$ \end{small}
\end{prop}

Propositions \ref{prop:complete} and \ref{prop:epsilonSound} are direct consequence of Lemma \ref{lemma:BLSH}.

\begin{remark}
	Proposition \ref{prop:epsilonSound} assumes that the Bloom filter hash functions are pseudo-random; this hypothesis is pretty standard for Bloom filter analysis. It can be achieved by using cryptographic hash functions with a random oracle-like behaviour.
\end{remark}

\begin{prop}[\textbf{Sender Privacy}] \label{prop:messPriv}
	Assume that the underlying cryptosystem is semantically secure and that the PIS function $\mathsf{update}_{BF}$ achieves User Privacy, then the scheme ensures Sender Privacy.
\end{prop}

\demo
If the scheme does not ensure Sender Privacy, that means that there exists an attacker who can distinguish between the output of $\mathsf{Send}(x_0, pk)$ and $\mathsf{Send}(x_1, pk)$, after the execution of $\mathsf{Send}(x_i, pk)$, $i\in\discretAB{2}{\Omega}$.

Note that the content of the Bloom filter buckets does not reveal information that can permit to distinguish between $x_0$ and $x_1$. Indeed, the only information $\mathcal{A}$ has with the filter structure is a set of $\Enc(\varphi(x_i))$ placed at different indexes $h^c(x_i)$, $i=e, 2, \ldots, \Omega$. Thanks to the semantic security of $\Enc$, this does not permit to distinguish between $\varphi(x_0)$ and $\varphi(x_1)$.

This implies that, with inputs $\Enc(x_i)$, $\mathsf{update_{BF}}(\Enc(\varphi(x_i)), h^c(x_i))$ ( for $i\geq 2$), the attacker can distinguish between $\Enc(x_0)$, $\mathsf{update_{BF}}(\Enc(\varphi(x_0)), h^c(x_0))$ and $\Enc(x_1)$,  $\mathsf{update_{BF}}(\Enc(\varphi(x_1)),h^c(x_1))$.

As $\mathsf{update_{BF}}$ does not leak information on its inputs, that means that the attacker can distinguish between $\Enc(x_0)$ and $\Enc(x_1)$ by choosing some other inputs to $\Enc$. That contradicts the semantic security assumption.
\findemo

\begin{prop}[\textbf{Receiver Privacy}] \label{prop:reqPriv}
	Assume that the PIR ensures User Privacy, then the scheme ensures Receiver Privacy.
\end{prop}

\demo

This property is a direct deduction of the PIR's User Privacy, as the only information $\Se$ gets from the execution of a $\mathsf{Retrieve}$ is a set of $\mathsf{Query}^{PIR}$. \findemo

These properties show that this protocol for Error-Tolerant Searchable Encryption has the security properties that we looked for. LSH functions are used in such a way that they do not degrade the security properties of the system.

\section{Application to Identification with Encrypted Biometric Data}\label{sec:ident}

\subsection{Our Biometric Identification System}

We now apply  our construction for error-tolerant searchable encryption to our biometric identification purpose. Thanks to the security properties of the above construction, this enables us to design a biometric identification system which achieves the security objectives stated in Section \ref{sec:informal}.

While applying the primitives of error-tolerant searchable encryption, the database $\mathcal{DB}$ 
takes the place of the server $\mathcal{S}$; the role of the Identity Provider $\mathcal{IP}$ varies with the step we are involved in. During the Enrolment step, $\mathcal{IP}$ behaves as  $\mathcal{X}$, and as $\mathcal{Y}$ during the Identification step. In this step, $\mathcal{IP}$ is in possession of the private key $sk$ used for the $\mathsf{Retrieve}$ query.

\subsubsection{Enrolment} 
\begin{itemize}
\item To enrol a user $\mathcal{U}_i$, the sensor $\mathcal{SC}$ acquires a sample $b_i$ from his biometrics and sends it to $\mathcal{IP}$,
\item The Identity Provider $\mathcal{IP}$ then executes  $\mathsf{Send}_{\X, \Se}(b_i, pk)$.
\end{itemize}

%After, for the Identification step, $\mathcal{IP}$ is  $\mathcal{Y}$ and is in possession of the private key. 

\subsubsection{Identification} \label{sec:subId}
\begin{itemize}
\item $\mathcal{SC}$ captures a fresh biometric template $b'$ from a user $\mathcal{U}$ and sends it to $\mathcal{IP}$,.
	\item The Identity Provider $\mathcal{IP}$ then executes  $\mathsf{Retrieve}_{\Y, \Se}(b', sk)$.
\end{itemize}

	At the end of the identification, $\mathcal{IP}$ has the fresh biometric template $b'$ along with the address of the candidate reference templates in $\DB$. To reduce the list of identities, we can use a secure matching scheme \cite{DBLP:conf/cans/BringerCPT07,SchoenmakersT06} to run a final secure comparison between $b'$ and the candidates. %This allows to reduce the list of identities. 
%	
%
%
%Note that this identification scheme, made of the enrolment and identification steps, complies with the security conditions enunciated in Sec. \ref{sec:model}. For space limitation reasons, the formal proof of this assertion is not given in this paper.

\subsection{Practical Considerations} \label{sec:pract}

 \subsubsection{\noindent {\bf Choosing the LSH family: an Example}}
 
 Let's place ourself in the practical setting of human identification through iris recognition. A well-known method to doing so is to use Daugman's IrisCode \cite{Daugman93}. This extracts a 2048-bit vector, along with a ``mask'', that defines the relevant information in this vector. Iris recognition is then performed by computing a simple Hamming distance; vectors that are at a Hamming distance less than a given threshold are believed to come from the same individual, while vectors that come from different eyes will be at a significantly larger distance.
 
 %To be more specific, Hao~\emph{et al.} \cite{HaAnDa06} report statistics that clearly show the separation between the matching scores and the non-matching ones. In particular, we can model the non-matching curve by a binomial distribution of about 250 independent trials with a success mean of $0.498$, as it is done in \cite{Daugma03}. On the other hand, the matching curve roughly behaves like a normal distribution of mean $0.034$ and of standard deviation $0.038$. 

 There are several paths to design LSH functions adapted to this kind of data. Random projections such as those defined in \cite{KushilevitzOR98}, is a convenient way to create LSH functions for binary vectors. However, for the sake of simplicity, we propose to use the functions used in \cite{Hao07}, in which they are referred as 'beacon indexes'. These functions are based on the fact that all IrisCode bits do not have the same distribution probability.
 
 In a few words, these functions first reorder the bits of the IrisCode by rows, so that in each row, the bits that are the most likely to induce an error are the least significant ones. The column are then reordered to avoid correlations between following bits. The most significant bits of rows are then taken as 10-bit hashes.
  The efficiency of this approach is demonstrated in \cite{Hao07} where the authors apply these LSH functions to identify a person thanks to his IrisCode. They interact with the UAE database which contains $N=632 500$ records; trivial identification would then require about $N/2$ classical matching computation, which is way too much for a large database. Instead, they apply $\mu = 128$ of those hashes to the biometric data, and look for IrisCodes that get the same LSH results for at least 3 functions. In doing this, they limit the number of necessary matching to $41$ instead of $N$.
  
 To determine the LSH capacity of these hash functions is not easy to do with real data; however, if we model $b$ and $b'$ as binary vectors such that the each bit of $b$ is flipped with a fixed probability (\textit{i.e.} if $b'$ is obtained out of $b$ through a binary symmetric channel), then the family induced is $(r_1, r_2, 1-(1-\frac{r_1}{2048})^{10}, (1-\frac{r_2}{2048})^{10})$-LSH. This estimation is conservative as IrisCodes are designed for biometric matching.

Combining these functions with a Bloom filter with storage in the way described in Sec. \ref{sec:Blsh} enables to have an secure identification scheme.

 \subsubsection{\noindent {\bf Overall complexity and efficiency}}
	
	We here evaluate the computational complexity of an identification request on the client's side. 
	We note $\kappa(op)$ the cost of operation $op$, and $|S|$ the size of the set $S$.
	Recalling Section \ref{sec:detail}, %.~\ref{sec:subId}, 
	the overall cost of a request is:
	\begin{small}
	\begin{eqnarray*}
	\lefteqn{\kappa(\text{request})} \\ & = & |H^c| (\kappa(\text{hash}) + \kappa(PIR) + |T| \kappa(\Dec)) + \kappa(\text{intersection}) \\
	& \leq & |H^c| \left(\kappa\left(\text{h}_{BF}\right) + \kappa\left(\text{h}_{LSH}\right) + \kappa\left(PIR\right) + |T| \kappa\left(\Dec\right)\right)  + O(|T| |H^c|) \\
	\end{eqnarray*}
\end{small}
	We here used data structures in which intersection of sets is linear in the set length, hence the term $O(|T| |H^c|)$; $|T|$ is the maximum size of a Bloom filter with storage bucket.
	
	To conclude this complexity estimation, let us recall that the cost of a hash function can be neglected in front of the cost of a decryption step. The PIR query complexity at the sensor level depends on the scheme used (recall that the PIR query is made only over the set of buckets and not over the whole database); in the case of Lipmaa's PIR \cite{Lipmaa05}, this cost $\kappa(PIR)$ is dominated by the cost of a Damg{\aa}rd-Jurik encryption. The overall sensor complexity of an identification request is $O(\mu \nu (|T| \kappa(\Dec) + \kappa(PIR)))$. 
	
	\section{Conclusion} \label{sec:ccl}
	
	This paper details the first non-trivial construction for biometric identification over encrypted binary templates. This construction meets the privacy model one can expect from an identification scheme and the computation costs are sublinear in the size of the database.
		
	We  studied identification scheme using binary data, together with Hamming distance. We plan to extend our scope to other metrics. 	
	A first lead to follow is to use techniques from \cite{KushilevitzOR98} which reduce the problem of \highlight{ANN} over Euclidean spaces into \highlight{ANN} over a Hamming space.

%%%%%%%%%%%%%%%%%%%%%%%%%%%%%%%%%%%%%%%%%%%%%%%%%%%%%%%%%%%%%%%%%%%%%%%%%%%%%%%%%%%%%%%%%%%%%%%%%%%%%%%%%%%%
%\bibliographystyle{plain}
%\bibliography{biosearch}

\begin{thebibliography}{10}

\bibitem{AndoniI08}
A.~Andoni and P.~Indyk.
\newblock Near-optimal hashing algorithms for approximate nearest neighbor in
  high dimensions.
\newblock {\em Commun. ACM}, 51(1):117--122, 2008.

\bibitem{BethencourtSW06}
J.~Bethencourt, D.~X. Song, and B.~Waters.
\newblock New constructions and practical applications for private stream
  searching (extended abstract).
\newblock In {\em IEEE Symposium on Security and Privacy}, pages 132--139. IEEE
  Computer Society, 2006.

\bibitem{Bloom70}
B.~H. Bloom.
\newblock Space/time trade-offs in hash coding with allowable errors.
\newblock {\em Commun. ACM}, 13(7):422--426, 1970.

\bibitem{BoCoRa02}
Ruud~M. Bolle, Jonathan~H. Connell, and Nalini~K. Ratha.
\newblock Biometric perils and patches.
\newblock {\em Pattern Recognition}, 35(12):2727--2738, 2002.

\bibitem{BonehCOP04}
D.~Boneh, G.~Di~Crescenzo, R.~Ostrovsky, and G.~Persiano.
\newblock Public key encryption with keyword search.
\newblock In Cachin and Camenisch \cite{DBLP:conf/eurocrypt/2004}, pages
  506--522.

\bibitem{BonehKOS07}
D.~Boneh, E.~Kushilevitz, R.~Ostrovsky, and W.~E. Skeith~III.
\newblock Public key encryption that allows {PIR} queries.
\newblock In {\em CRYPTO}, volume 4622, pages 50--67. Springer, 2007.

\bibitem{DBLP:conf/africacrypt/BringerC08}
J.~Bringer and H.~Chabanne.
\newblock An authentication protocol with encrypted biometric data.
\newblock In Serge Vaudenay, editor, {\em AFRICACRYPT}, volume 5023 of {\em
  Lecture Notes in Computer Science}, pages 109--124. Springer, 2008.

\bibitem{BringCCKZ07}
J.~Bringer, H.~Chabanne, G.~Cohen, B.~Kindarji, and G.~Zémor.
\newblock Theoretical and practical boundaries of binary secure sketches.
\newblock {\em IEEE Transactions on Information Forensics and Security},
  3(4):673--683, 2008.

\bibitem{BrChDo06}
J.~Bringer, H.~Chabanne, and Q.~D. Do.
\newblock A fuzzy sketch with trapdoor.
\newblock {\em IEEE Transactions on Information Theory}, 52(5):2266--2269,
  2006.

\bibitem{DBLP:conf/acisp/BringerCIPTZ07}
J.~Bringer, H.~Chabanne, M.~Izabach{\`e}ne, D.~Pointcheval, Q.~Tang, and
  S.~Zimmer.
\newblock An application of the {Goldwasser-Micali} cryptosystem to biometric
  authentication.
\newblock In Josef Pieprzyk, Hossein Ghodosi, and Ed~Dawson, editors, {\em
  ACISP}, volume 4586 of {\em Lecture Notes in Computer Science}, pages
  96--106. Springer, 2007.

\bibitem{BoBW07}
J.~Bringer, H.~Chabanne, and B.~Kindarji.
\newblock The best of both worlds: Applying secure sketches to cancelable
  biometrics.
\newblock {\em Science of Computer Programming}, 74(1--2):43--51, 2008.
\newblock Special Issue on Security and Trust.

\bibitem{DBLP:conf/cans/BringerCPT07}
J.~Bringer, H.~Chabanne, D.~Pointcheval, and Q.~Tang.
\newblock Extended private information retrieval and its application in
  biometrics authentications.
\newblock In Feng Bao, San Ling, Tatsuaki Okamoto, Huaxiong Wang, and Chaoping
  Xing, editors, {\em CANS}, volume 4856 of {\em Lecture Notes in Computer
  Science}, pages 175--193. Springer, 2007.

\bibitem{DBLP:conf/iwsec/BringerCPZ08}
J.~Bringer, H.~Chabanne, D.~Pointcheval, and S.~Zimmer.
\newblock An application of the boneh and shacham group signature scheme to
  biometric authentication.
\newblock In Kanta Matsuura and Eiichiro Fujisaki, editors, {\em IWSEC}, volume
  5312 of {\em Lecture Notes in Computer Science}, pages 219--230. Springer,
  2008.

\bibitem{BrCT07}
J.~Bringer, H.~Chabanne, and Q.~Tang.
\newblock An application of the {Naccache-Stern} knapsack cryptosystem to
  biometric authentication.
\newblock In {\em AutoID}, pages 180--185. IEEE, 2007.

\bibitem{ByunLL06}
J.~W. Byun, D.~H. Lee, and J.~Lim.
\newblock Efficient conjunctive keyword search on encrypted data storage
  system.
\newblock In Andrea~S. Atzeni and Antonio Lioy, editors, {\em EuroPKI}, volume
  4043, pages 184--196. Springer, 2006.

\bibitem{DBLP:conf/eurocrypt/2004}
C.~Cachin and J.~Camenisch, editors.
\newblock {\em Advances in Cryptology - EUROCRYPT 2004, International
  Conference on the Theory and Applications of Cryptographic Techniques,
  Interlaken, Switzerland, May 2-6, 2004, Proceedings}, volume 3027 of {\em
  LCNS}. Springer, 2004.

\bibitem{ChorKGS98}
B.~Chor, E.~Kushilevitz, O.~Goldreich, and M.~Sudan.
\newblock Private information retrieval.
\newblock {\em J. ACM}, 45(6):965--981, 1998.

\bibitem{Daugman93}
J.~Daugman.
\newblock High confidence visual recognition of persons by a test of
  statistical independence.
\newblock {\em IEEE Trans. Pattern Anal. Mach. Intell.}, 15(11):1148--1161,
  1993.

\bibitem{DodisRS04}
Y.~Dodis, L.~Reyzin, and A.~Smith.
\newblock Fuzzy extractors: How to generate strong keys from biometrics and
  other noisy data.
\newblock In Cachin and Camenisch \cite{DBLP:conf/eurocrypt/2004}, pages
  523--540.

\bibitem{Gamal84}
T.~El~Gamal.
\newblock A public key cryptosystem and a signature scheme based on discrete
  logarithms.
\newblock In {\em CRYPTO}, pages 10--18, 1984.

\bibitem{gasarch:1}
W.~I. Gasarch.
\newblock A survey on private information retrieval.
\newblock http://www.cs.umd.edu/~gasarch/pir/pir.html.

\bibitem{GentryR05}
C.~Gentry and Z.~Ramzan.
\newblock Single-database private information retrieval with constant
  communication rate.
\newblock In {\em ICALP}, volume 3580 of {\em LCNS}, pages 803--815. Springer,
  2005.

\bibitem{GertnerIKM98}
Y.~Gertner, Y.~Ishai, E.~Kushilevitz, and T.~Malkin.
\newblock Protecting data privacy in private information retrieval schemes.
\newblock In {\em STOC}, pages 151--160, 1998.

\bibitem{goh03}
E.-J. Goh.
\newblock Secure indexes.
\newblock Cryptology ePrint Archive, Report 2003/216, 2003.
\newblock \url{http://eprint.iacr.org/2003/216/}.

\bibitem{GoldwasserM84}
S.~Goldwasser and S.~Micali.
\newblock Probabilistic encryption.
\newblock {\em J. Comput. Syst. Sci.}, 28(2):270--299, 1984.

\bibitem{Hao07}
F.~Hao, J.~Daugman, and P.~Zielinski.
\newblock A fast search algorithm for a large fuzzy database.
\newblock {\em IEEE Trans. on Inf. Forensics and Security}, 2008.

\bibitem{HaAnDa06}
Feng Hao, R.~Anderson, and J.~Daugman.
\newblock Combining crypto with biometrics effectively.
\newblock {\em Computers, IEEE Transactions on}, 55(9):1081--1088, Sept. 2006.

\bibitem{HerzbergJKY95}
A.~Herzberg, S.~Jarecki, H.~Krawczyk, and M.~Yung.
\newblock Proactive secret sharing or: How to cope with perpetual leakage.
\newblock In {\em CRYPTO}, volume 963, pages 339--352. Springer, 1995.

\bibitem{indyk04nearest}
P.~Indyk.
\newblock Nearest neighbors in high-dimensional spaces.
\newblock In {\em Handbook of Discrete and Computational Geometry, chapter 39}.
  CRC Press, 2004.
\newblock 2rd edition.

\bibitem{IndykM98}
P.~Indyk and R.~Motwani.
\newblock Approximate nearest neighbors: Towards removing the curse of
  dimensionality.
\newblock In {\em Symposium on the Theory Of Computing}, pages 604--613, 1998.

\bibitem{JaNaNa08}
A.~K. Jain, K.~Nandakumar, and A.~Nagar.
\newblock Biometric template security.
\newblock {\em EURASIP Journal on Advances in Signal Processing,}, Special
  Issue on Advanced Signal Processing and Pattern Recognition Methods for
  Biometrics, 2008.

\bibitem{JainPHP99}
A.~K. Jain, S.~Prabhakar, L.~Hong, and S.~Pankanti.
\newblock Fingercode: A filterbank for fingerprint representation and matching.
\newblock In {\em CVPR}, pages 2187--. IEEE Computer Society, 1999.

\bibitem{JaRoUl05}
A.~K. Jain, A.~Ross, and U.~Uludag.
\newblock Biometric template security: Challenges and solutions.
\newblock In {\em Proc. of 13th European Signal Processing Conference
  (EUSIPCO)}, Antalya, Turkey, September 2005.

\bibitem{JuelsW99}
A.~Juels and M.~Wattenberg.
\newblock A fuzzy commitment scheme.
\newblock In {\em ACM Conference on Computer and Communications Security},
  pages 28--36, 1999.

\bibitem{Khader06}
D.~Khader.
\newblock Public key encryption with keyword search based on {K}-resilient
  {IBE}.
\newblock In {\em ICCSA (3)}, volume 3982, pages 298--308. Springer, 2006.

\bibitem{KirschMitzen06}
A.~Kirsch and M.~Mitzenmacher.
\newblock Distance-sensitive {B}loom filters.
\newblock In {\em Algorithm Engineering \& Experiments}, Jan 2006.

\bibitem{KushilevitzOR98}
E.~Kushilevitz, R.~Ostrovsky, and Y.~Rabani.
\newblock Efficient search for approximate nearest neighbor in high dimensional
  spaces.
\newblock In {\em Symposium on the Theory Of Computing}, pages 614--623, 1998.

\bibitem{Lipmaa05}
H.~Lipmaa.
\newblock An oblivious transfer protocol with log-squared communication.
\newblock In {\em ISC}, volume 3650, pages 314--328. Springer, 2005.

\bibitem{NaNaJa07}
Karthik Nandakumar, Abhishek Nagar, and Anil~K. Jain.
\newblock Hardening fingerprint fuzzy vault using password.
\newblock In Seong-Whan Lee and Stan~Z. Li, editors, {\em ICB}, volume 4642 of
  {\em Lecture Notes in Computer Science}, pages 927--937. Springer, 2007.

\bibitem{OstrovskyS97}
R.~Ostrovsky and V.~Shoup.
\newblock Private information storage (extended abstract).
\newblock In {\em STOC}, pages 294--303, 1997.

\bibitem{OstSke07a}
R.~Ostrovsky and W.~E. Skeith~III.
\newblock Algebraic lower bounds for computing on encrypted data.
\newblock Cryptology ePrint Archive, Report 2007/064, 2007.
\newblock \url{http://eprint.iacr.org/}.

\bibitem{Pa99}
P.~Paillier.
\newblock Public-key cryptosystems based on composite degree residuosity
  classes.
\newblock In {\em Advances in Cryptology, Proceedings of EUROCRYPT '99}, volume
  1592 of {\em LCNS}, pages 223--238. Springer, 1999.

\bibitem{RyuT07}
E.-K. Ryu and T.~Takagi.
\newblock Efficient conjunctive keyword-searchable encryption.
\newblock In {\em AINA Workshops (1)}, pages 409--414. IEEE Computer Society,
  2007.

\bibitem{SchoenmakersT06}
B.~Schoenmakers and P.~Tuyls.
\newblock Efficient binary conversion for {P}aillier encrypted values.
\newblock In {\em EUROCRYPT}, volume 4004, pages 522--537. Springer, 2006.

\bibitem{Shamir79}
A.~Shamir.
\newblock How to share a secret.
\newblock {\em Commun. ACM}, 22(11):612--613, 1979.

\bibitem{SuLiMe07}
Yagiz Sutcu, Qiming Li, and N.~Memon.
\newblock Protecting biometric templates with sketch: Theory and practice.
\newblock {\em Information Forensics and Security, IEEE Transactions on},
  2(3):503--512, Sept. 2007.

\bibitem{DBLP:conf/ispec/TangBCP08}
Q.~Tang, J.~Bringer, H.~Chabanne, and D.~Pointcheval.
\newblock A formal study of the privacy concerns in biometric-based remote
  authentication schemes.
\newblock In Liqun Chen, Yi~Mu, and Willy Susilo, editors, {\em ISPEC}, volume
  4991 of {\em Lecture Notes in Computer Science}, pages 56--70. Springer,
  2008.

\bibitem{TuylsAKSBV05}
P.~Tuyls, A.~H.~M. Akkermans, T.~A.~M. Kevenaar, G.~Jan Schrijen, A.~M. Bazen,
  and R.~N.~J. Veldhuis.
\newblock Practical biometric authentication with template protection.
\newblock In {\em Audio-and Video-Based Biometrie Person Authentication},
  volume 3546, pages 436--446. Springer, 2005.

\end{thebibliography}

%%%%%%%%%%%%%%%%%%%%%%%%%%%%%%%%%%%%%%%%%%%%%%%%%%%%%%%%%%%%%%%%%%%%%%%%%%%%%%%%%%%%%%%%%%%%%%%%%%%%%%%%%%%%
\appendix

\section{Achieving Symmetric Receiver Privacy} \label{sec:achievingSymReqPriv}

In this section, we introduce a new security concern, which we call \emph{Symmetric Receiver Privacy}. 

\subsection{Condition statement} \label{sec:symCondState}

This property aims at limiting the amount of information that $\Y$ gets through the protocol. Indeed, if previous constructions of Searchable Encryption such as \cite{BonehCOP04,goh03} seem to consider that the sender and the receiver are the same person, thus owning the database in the same way, there are applications where the receiver must not dispose of the entire database. 
If for example different users $\Y_i$ have access to the application, we do not want user $\Y_i$ to obtain information on another user $\Y_j$'s data.

For this purpose, we define a database simulator $\Se_1$.
$\Se_1(x')$ is a simulator which only knows the tags of the registered elements that are in $\Phi(x')$, while the other elements are random. Here, $x'$ stands for the message to be retrieved. On the other hand, $\Se_0$ is the regular server, which genuinely runs the protocol.

\begin{cond}[Symmetric Receiver Privacy]\label{cond:symReqPriv} \label{cond:symRecPriv}
The scheme is said to respect \emph{Symmetric Receiver Privacy} if there exists a simulator $\Se_1$ such that the advantage of any malicious receiver is negligible in the $\mathsf{Exp}_{\mathcal{A}}^{\text{Sym-Rec-Privacy}}$ experiment described below. Here, $\mathcal{A}$ is the 'honest-but-curious' opponent taking the place of $\Y$, and $\mathcal{C}$ the challenger at the server side.
	\begin{small}
	$$
		\begin{array}{ll}
				\mathsf{Exp}_{\mathcal{A}}^{\text{Sym-Rec-Privacy}} & \\
			& \hspace{-2cm}\vline
				\begin{array}{clclc}
					1.	& (pk,sk) & \lar & \mathsf{KeyGen}(1^k)			&(\mathcal{A})	\\
					%2.	& \{x_1, \ldots, x_\Omega\}, d(x_i, x_j)> \lambda_{max}, \forall i,j\in\discretAB{1}{\Omega}	& \lar	& \mathcal{A}	 & (\mathcal{A})\\
					2.	& \{x_1, \ldots, x_\Omega\}, 	& \lar	& \mathcal{A}	 & (\mathcal{A})\\
					& \multicolumn{3}{l}{d(x_i, x_j)> \lambda_{max}, \forall i,j\in\discretAB{1}{\Omega}} \\
					3.	& \varphi(x_i) & \lar & \mathsf{Send}_{\mathcal{A}, \Se}(x_i, pk)	& (\mathcal{A})	\\
					4. & e\in_R\{0,1\} 								& \lar	& \mathcal{C} & (\mathcal{C})\\
					%%5. & \{x'_1, \ldots, x'_p\}, d(x'_i, x'_j)> \lambda_{max}, \forall i,j\in\discretAB{1}{p}					& \lar	& \mathcal{A}	& (\mathcal{A})\\
					5. & \{x'_1, \ldots, x'_p\}, 					& \lar	& \mathcal{A}	& (\mathcal{A})\\
					& \multicolumn{3}{l}{d(x'_i, x'_j)> \lambda_{max}, \forall i,j\in\discretAB{1}{p}} \\
					6. & \Phi(x'_i) & \lar & \mathsf{Retrieve}_{\mathcal{A}, \Se_e}(x'_i, sk) & (\mathcal{A})	\\
					7. & e' \in \{0,1\}									& \lar	& \mathcal{A} & (\mathcal{A})
				\end{array} 
		\end{array}
	$$
	\end{small}
The advantage of the adversary is $|\Pp{e'=e} - \frac{1}{2}|$.
\end{cond}

This new condition does not fit into previous models for Searchable Encryption, and is not satisfied by constructions such as \cite{BonehKOS07,goh03}. It is inspired by the Data Privacy property of SPIR protocols, which states that it is not possible to tell whether or not $\Se$ possesses more data than the received messages. 
Indeed, if the receiver is able to tell the difference between a server $\Se_0$ that possess more data than what $\Y$ received, and a server $\Se_1$ that just has in memory the information that $\Y$ needs, then $\Y$ detains more information than what he ought to; that is why this indistinguishability game fits the informal description of \emph{Symmetric Receiver Privacy}.

Section \ref{sec:symRecConstr} is dedicated to give a construction that also fits this security conditions.

\subsection{Specific Tools} 

\paragraph{ElGamal} For this purpose, we specify a second cryptosystem $\left ( \mathcal{S}etup, \mathcal{E}nc, \mathcal{D}ec \right )$ to be that of ElGamal: let $\mathcal{G}$ be a cyclic group of order $q$, a large prime, with $g$ a generator; let $f$ be another generator of $\mathcal{G}$. $\mathcal{S}etup$ renders the key pair $(h=g^v,v)$ for $v$ a random integer. Encryption $\mathcal{E}nc$ takes a random value $r$, and computes $\mathcal{E}nc(x) = (g^r, h^r x)$. The value $\mathcal{D}ec\left((y_1,y_2 )\right ) = \frac{y_2}{y_1^v} = x$ can be computed thanks to the secret key $v$. The homomorphic property is $\mathcal{D}ec(\mathcal{E}nc(x)  \mathcal{E}nc(x')) = x x'$.

\paragraph{Secret splitting} Let $s$ be a \emph{small} secret; we wish to split $s$ into $n$ re-randomizable parts. There is a general technique for this, called Proactive Secret Sharing \cite{Shamir79,HerzbergJKY95}, but for clarity reasons, we propose a simple technique for this. We construct $n$ shares $A_1, \ldots, A_n$ such that $A_i = g^{r_i}$ where $r_i$ is a random integer, for $i\in\discretAB{1}{n-1}$ and $A_n = g^{-\sum{r_i} + s}$, where $g$ is the generator of a group of \emph{large} prime order $q$. Recovering $s$ can be done by multiplying all the $A_i$, and then proceeding to an exhaustive search to compute the discrete logarithm of $g^s$ in basis $g$. Re-randomization of the parts $A_i$ can easily be done by choosing a random integer $t$, and replacing each $A_i$ by $A_i^t$. The generator for the discrete logarithm must then be replaced by $g^t$.

\subsection{Extending our Scheme} \label{sec:symRecConstr}

The scheme proposed in Sec. \ref{sec:detail} does not achieve Symmetric Receiver Privacy. For example, the user $\Y$ has access to all the $\varphi(x_i)$ such that there exists $h^c,h^c_0\in H^c, h^c(x_i)=h^c_0(x')$. Without further caution, a malicious user could get more information than what he ought to. We here describe an example of a protocol variant that leads to the desired properties.

We will apply secret splitting to the tags $\varphi(x)$ returned by $\mathsf{Send}$. That implies that we consider the range of $\varphi(x)$ to be relatively small, for example of $32$-bit long integers.
Primitives are adapted this way:
%\vspace{-0.2cm}
\begin{itemize}
	\item $\mathsf{KeyGen}(1^k)$ is unchanged, but here both $\mathsf{Setup}$ and $\mathcal{S}etup$ are used to generate $(pk,sk)$,
	\item $\mathsf{Send}_{\X, \Se}(x, pk)$ is slightly modified, namely:
		\begin{enumerate}
			\item {\bf Identifier establishment (\emph{unchanged})} $\Se$ attributes to $x$ a unique identifier $\varphi(x)$, and sends it to $\X$.
			\item {\bf Data storage (\emph{unchanged})} $\X$ sends $\Enc(x)$ to $\Se$, who stores it in a memory cell that depends on $\varphi(x)$.
			\item {\bf Data indexing}\begin{itemize}
				\item $\X$ splits the tag $\varphi(x)$ into $|H^c|$ shares $A_{x,1}, \ldots, A_{x,|H^c|}$ thanks to the method described above, and picks a random integer $r_x$,
				\item $\X$ computes all $h^c_i(x)$, 
			 and executes the queries $$\mathsf{update_{BF}}((\mathcal{E}nc(f^{r_x}),\mathcal{E}nc(A_{x,i})), h^c_i(x))$$ to send $(\mathcal{E}nc(f^{r_x}),\mathcal{E}nc(A_{x,i}))$ to be added to the filter's bucket of index $h^c_i(x)$, where  $h^c_i$ is the $i$-th function of $H^c$, for $i\in\discretAB{1}{|H^c|}$.
				\end{itemize}
		\end{enumerate}
		At the end of this update, the bucket $T_\alpha$ of the filter is filled with $l$ couples \\$(\mathcal{E}nc(f^{z_{\alpha,j}}),\mathcal{E}nc(B_{\alpha,j})), j\in\discretAB{1}{l}$. $B_{\alpha,j}$ is a share of some tag, or a random element of the group.
	\item $\mathsf{Retrieve}_{\Y, \Se}(x', sk)$ is adapted consequently:
		\begin{enumerate}
			\item {\bf (\emph{unchanged})} $\Y$ computes each $\alpha_i = h_i^c(x')$ for $h_i \in H^c$, then executes $\mathsf{Query}^{PIR}_{\Y, \Se}(\alpha_i)$,
			\item \begin{itemize}
				\item $\Se$ first re-randomizes the content of each bucket of the Bloom filter database by the same random value.  The filter bucket $T_{\alpha_i}=\left \{(\mathcal{E}nc(f^{z_{\alpha_i,j}}),\mathcal{E}nc(B_{\alpha_i,j})), j\in\discretAB{1}{l}\right \}$ becomes 
			%\vspace{-0.1cm} 
			\begin{small}
			$$\hspace{-1cm}T_{\alpha_i}^{c_1,c_2}=\left \{(\mathcal{E}nc(f^{z_{\alpha_i,j}})^{c_1},\mathcal{E}nc(B_{\alpha_i,j})^{c_2}), j\in\discretAB{1}{l}\right \}$$
			\end{small}
				\item $\Se$ then answers to the PIR Query, and sends along $g^{c_2}$ to $\Y$,
			\end{itemize}
%			\item $\Se$ responds to the $\mathsf{Query}$ after re-randomizing the content of each bucket of the Bloom filter database by the same random value. The filter bucket $T_{\alpha_i}=\left \{(\mathcal{E}nc(f^{z_{\alpha_i,j}}),\mathcal{E}nc(B_{\alpha_i,j})), j\in\discretAB{1}{l}\right \}$ becomes 
%			\vspace{-0.1cm} $$T_{\alpha_i}^{c_1,c_2}=\left \{(\mathcal{E}nc(f^{z_{\alpha_i,j}})^{c_1},\mathcal{E}nc(B_{\alpha_i,j})^{c_2}), j\in\discretAB{1}{l}\right \}$$
%			 \vspace{-0.1cm} $\Se$ also sends $g^{c_2}$ to $\Y$,
			\item $\Y$ decrypts the content of each bucket $T_{\alpha_i}^{c_1,c_2}$ to get a set of couples $(f^{z_{\alpha_i,j} c_1 }, B_{\alpha_i,j}^{c_2})$,
			\item If the same element $f^{z c_1}$ is present in the intersection of all the different sets $T_{\alpha_i}^{c_1,c_2}$, then $\Y$ possesses all shares of a tag $\varphi(x)$, and then computes $\prod_{i=1}^{|H^c|} A_{x,i}^{c_2} = \left (g^{c_2}\right )^{\varphi(x)}$,
			\item $\Y$ finally runs a discrete logarithm of $\left (g^{c_2}\right )^{\varphi(x)}$ in basis $g^{c_2}$, and adds $\varphi(x)$ to the set of results $\Phi(x')$.
		\end{enumerate}
\end{itemize}%\vspace{-0.2cm}

Note that this scheme can also be generalized for other Proactive Secret Sharing schemes.

\subsection{Security Properties}

This new scheme is an extension of the previous one, and the same security properties are achieved. Moreover, Condition \ref{cond:symReqPriv} also holds.
Indeed, the modification to the $\mathsf{Send}$ procedure is not significant enough to alter the Sender Privacy property: the only modification on $\Se$'s side is the content of the $\mathsf{update_{BF}}$ procedure, which does not leak. Moreover, the Receiver Privacy property is also preserved, as communications from $\Y$ to $\Se$ in $\mathsf{Retrieve}$ only involves a PIR query.

\begin{prop}[Symmetric Receiver Privacy] \label{prop:SymReqPriv}
Assume the PIR ensures Data Privacy \textit{i.e.} it is a SPIR, and that $H$ is a $(\lambda_{min}, \lambda_{max}, \epsilon_1, \epsilon_2)$-LSH family with a negligible $\epsilon_2$, then the scheme ensures Symmetric Receiver Privacy, over the Decisional Diffie Hellman  hypothesis.
\end{prop}

To demonstrate this proposition, let us begin with a preliminary Lemma.
\begin{lemma}
	Let $s_1, \ldots, s_t \in S$ be $t$ different secrets, with $|S|$ small. %TODO : ecrire correctement ce que c'est que small.
	 Let $A_{i,1}, \ldots, A_{i,n}$ be the $n$ parts of the secret $s_i$ split thanks to the method described in Sec. \ref{sec:achievingSymReqPriv}. Let $\pi_0 \subset \{A^{c}_{i,j}, i\in\discretAB{1}{t}, j\in \discretAB{1}{n}, c\in\discretAB{1}{q}\}$ be collection of $k$ such parts, and $\pi_1 = \{g^{r_1}, \ldots, g^{r_k}\}$ a set of $k$ random elements of the cyclic group $\mathcal{G}$.
	
	Under the DDH assumption, if an adversary $\mathcal{A}$ can distinguish between $\pi_0$ and $\pi_1$, then there exists $c_0\in \discretAB{1}{q}, i\in\discretAB{1}{t}$ such that $\{A_{i,1}^{c_0}, \ldots, A_{i,n}^{c_0}\} \subset \pi_0$.
\end{lemma}

{ \em Sketch of proof}

Let $(g, g^a, g^b, g^c)$ be an instance of the DDH problem. An adversary can solve this instance if he can tell, with non-negligible probability, whether $g^c = g^{ab}$ or not.

We take $t=1$, because all secrets are independent, and $n=2$ (it is easy to take $n>2$ by multiplying the parts and returning to the case $n=2$). Suppose the lemma is false, that means that there exists a polynomial algorithm $\mathcal{A}$ that takes as inputs couples $(g^{c_u}, g^{c_u  r})$ and $(g^{c_v}, g^{c_v (s-r)})$, with $c_u\neq c_v$, and that returns the secret $s$ with non-negligible probability.

We then give as input to $\mathcal{A}$ the couples $(g, g^a)$ and $(g^b, g^{bs-c})$ for $s\in S$. If $\mathcal{A}$ returns $s$, that means that $g^{c-bs} = g^{b  (a-s)} =g^{ba}g^{-bs}$. We %can run $\mathcal{A}$ on each $s\in S$, and 
 finally have an advantage on the DDH problem; that proves the lemma.
\findemo

\vspace{0.2cm}
{ \em Proof of Proposition \ref{prop:SymReqPriv}}

We now build a simulator $\Se_1$ for the server in order to prove the proposition. Let $x'$ be the request and $\Phi(x')=\{\varphi(x_{1}), \ldots, \varphi(x_{k})\}$ be the genuine answer to $\mathsf{Retrieve}(x',sk)$. First, the simulator generates $\Omega$ random elements $\{z_1, \ldots, z_\Omega\} \subset \discretAB{1}{q}$; he associates the first $k$ elements to the elements of $\Phi(x')$.
The simulator splits each of the $\varphi(x_j)$ into the $n=|H^c|$ parts $A_{x_j,1}, \ldots, A_{x_j,n}$. Finally, he picks random integers $c_1, c_2$.

Since the PIR is symmetrical, we can impose the response to each $\mathsf{Query}(\alpha_i)$ to be a set containing the $k$ elements that must be present in the intersection, namely $\left ( \mathcal{E}nc(f^{z_j})^{c_1}, \mathcal{E}nc(A_{x_j,i})^{c_2} \right )$, and the remaining random values $$\left( \mathcal{E}nc(f^z)^{c_1}, \mathcal{E}nc(g^r) \right ),$$ with $z$ a random element of $\{z_{k+1}, \ldots, z_\Omega\}$ and $r$ a random integer. We give to the simulator enough memory to remember which $z$ was returned for which $\alpha$, so that multiple queries to the same $\alpha$ are consistent. The simulator also returns $g^{c_2}$.

Let $\mathcal{A}$ be a malicious receiver in the $\mathsf{Exp}_\mathcal{A}^{\text{Sym-Rec-Privacy}}$ experiment. Following Cond.~\ref{cond:symReqPriv}, $\mathcal{A}$ makes $p$ $\mathsf{Retrieve}$ queries to $\Se$; each of these requests lead to $|H^c|$ calls to $\mathsf{Query}$. As the requests $x_i'$ are $\lambda_{max}$-separated, and as the hashes are $\lambda_{min}, \lambda_{max}, \epsilon_1,\epsilon_2$ with a negligible $\epsilon_2$, we can consider these $\mathsf{Retrieve}$ queries to be independent.

Note that the first parts of the Bloom filters are always indistinguishable, as they are generated in the same way. Therefore, if $\mathcal{A}$ distinguishes between $\Se_0$ and $\Se_1$, that means that he distinguished between a given $\pi_0$ and $\pi_1$, constructed by taking the set of all answers to the $\mathsf{Query}$ request he made. By application of the Lemma, we deduce the proposition.

\findemo

\end{document}